\documentclass[final,5p,times,twocolumn]{elsarticle}

\usepackage{amssymb}

\usepackage{xspace}	
\usepackage{color}

\newcommand{\sqrts}[1]{\mbox{$\sqrt{s_{_{NN}}}$}~=~#1~GeV\xspace} 

\journal{Physics Letter B}

\begin{document}

\begin{frontmatter}




\title{Conserved number fluctuations in a hadron resonance gas model}


\author[bhu]{
P.~Garg
}
\author[barc]{
D.~K.~Mishra
}
\author[barc]{
P.~K.~Netrakanti
}
\author[niser]{
B.~Mohanty
} 
\ead{bedanga@rcf.rhic.bnl.gov}
\author[barc]{
A.~K.~Mohanty
}
\author[bhu]{
B.~K.~Singh
}
\author[lbl,ccnu]{
N.~Xu
}


\address[bhu]{ Department of Physics, Banaras Hindu University, Varanasi 221005, India}
\address[barc]{ Nuclear Physics Division, Bhabha Atomic Research Center, Mumbai 400085, India}
\address[niser]{ School of Physical Sciences, National Institute of Science Education and Research, Bhubaneswar 751005, India}
\address[lbl]{Nuclear Science Division, Lawrence Berkeley National Laboratory, Berkeley, CA 94720, USA}
\address[ccnu]{Key Laboratory of the Ministry of Education of China, Central China Normal University, Wuhan, 430079, China}
\begin{abstract}
Net-baryon, net-charge and net-strangeness number fluctuations in high energy heavy-ion collisions
are discussed within the framework of a hadron resonance gas (HRG) model. Ratios of the conserved 
number susceptibilities calculated in HRG are being compared to the corresponding experimental 
measurements to extract information about the freeze-out condition and the phase structure of systems 
with strong interactions. We emphasize the importance of considering the actual experimental acceptances 
in terms of kinematics (pseudorapidity ($\eta$) and transverse momentum ($p_{T}$)), the detected 
charge state, effect of collective motion of particles in the system and the resonance decay 
contributions before comparisons are made to the theoretical calculations. 
In this work, based on HRG model, we report that the net-baryon number fluctuations are least 
affected by experimental acceptances compared to the net-charge and net-strangeness number fluctuations. 
\end{abstract}


\end{frontmatter}


\section{Introduction}
Measurement of the  moments of distribution for conserved quantities like net-baryon, net-charge 
and net-strangeness number for systems undergoing strong interactions as in high energy heavy-ion 
collisions, have recently provided rich physics insights 
~\cite{Gupta:2011wh,Gavai:2010zn,Karsch:2010ck,Bazavov:2012vg,Aggarwal:2010wy,Stephanov:2011pb,
BraunMunzinger:2011dn,Friman:2011pf,Yang:2011nj,Asakawa:2009aj}. The most crucial realization is 
that, the product of moments of the conserved number distributions are measurable experimentally 
and can be linked to susceptibilities ($\chi$) computed in Quantum Chromodynamic (QCD) based 
calculations~\cite{Gupta:2011wh,Aggarwal:2010wy}. For example, 
${\it S}$$\sigma$ = $\chi^{(3)}/\chi^{(2)}$ and
$\kappa$$\sigma^2$ = $\chi^{(4)}/\chi^{(2)}$, 
where $\sigma$ is the standard deviation, ${\it S}$ is the skewness, $\kappa$ is the kurtosis of 
the measured conserved number distribution, $\chi^{(n)}$ are the $n^{th}$ order theoretically calculated
susceptibilities associated with these conserved numbers. 
Such a connection between theory
and high energy heavy-ion collision experiment has led to furthering our understanding about the 
freeze-out conditions~\cite{Gavai:2010zn,Bazavov:2012vg}, details of the quark-hadron 
transition~\cite{Gupta:2011wh,Friman:2011pf} and plays a crucial role for the search of possible QCD 
critical point in the QCD phase diagram~\cite{Aggarwal:2010wy}. In all such physics cases there is a 
need to establish a reference point for the measurements. Computing these quantities within the 
framework of a hadron resonance gas (HRG) model~\cite{BraunMunzinger:2003zd} provides such a 
reference for both experimental data and QCD based calculations. 

The experimental measurements have limitations, they are usually available for a fraction
of the total kinematic phase space, due to the finite detector geometries and can detect only 
certain species of the produced particles. For example, measurements related 
to net-baryon number distribution is restricted by the kinematic range in $p_{T}$ where their 
identification is possible. In addition, baryons like neutron are not commonly measured in most of 
the high energy heavy-ion collision experiments. While for the net-charge number distribution, 
the limitation is usually in kinematic range available in $\eta$ and the details of how contribution 
from different charge states and resonances are dealt with in the measurements. The kinematic 
acceptance in a typical high energy heavy-ion collision experiment like STAR~\cite{McDonald:2012ts} 
and PHENIX~\cite{Mitchell:2012mx} at the Relativistic Heavy-Ion Collider facility (RHIC) for 
net-charge multiplicity distributions are: $|\eta|$ $<$ 0.5, 0.2 $<$ $p_{T}$ $<$ 2.0 GeV/$c$ and 
$|\eta|$ $<$ 0.35, 0.3 $<$ $p_{T}$ $<$ 1.0 GeV/$c$, respectively. While for net-baryon number and 
net-strangeness related studies carried out in the STAR experiment, within $|\eta|$ $<$ 0.5, is 
through the measurement of net-protons and net-kaons in the range of 0.4 $<$ $p_{T}$ $<$ 0.8 GeV/$c$ 
and 0.2 $<$ $p_{T}$ $<$ 2.0 GeV/$c$, respectively~\cite{Aggarwal:2010wy,McDonald:2012ts}. 

The main goal of this paper is to demonstrate using the HRG model (discussed in next section), the 
effect of the above experimental limitations on the physics observables  $\chi^{(3)}/\chi^{(2)}$ and
$\chi^{(4)}/\chi^{(2)}$. Our model based study clearly shows that the value of the observables
related to net-charge and net-strangeness strongly depends on the experimental kinematic and charge 
state acceptances. Where charge state could be electric charge ($Q$) = 1 or higher for net-charge 
measurements and strangeness number ($S$) = 1 or higher for net-strangeness measurements. In 
contrast, the net-baryon number studies are found to be minimally affected by these 
experimental limitations. In this work, we have not considered the initial baryon distribution due to the 
participant number fluctuations in the heavy-ion collisions on the results for net-baryon 
fluctuations~\cite{kon}. Another important effect that could influence the values of the
higher moments of the net-charge, net-strangeness and net-baryon number distributions in limited 
acceptance, are the conservation laws related to charge, strangeness and baryon number.

The paper is organized as follows. In the Section~\ref{sec:hrg}, we will discuss the HRG model used 
in this study. In the Section~\ref{sec:results}, the results for the observable 
$\chi^{(3)}/\chi^{(2)}$ and $\chi^{(4)}/\chi^{(2)}$ are presented for different kinematic 
acceptances, charge states,  effect of collective flow of particle in the system and the resonance 
decay contributions. We also provide a table listing the values of these observable for typical 
experimental conditions as encountered in STAR and PHENIX experiments at RHIC and ALICE experiment 
at the Large Hadron Collider (LHC) Facility. Finally in Section~\ref{sec:summary}, we summarize our 
findings and mention about the implications of this work to the current experimental measurements in 
high energy heavy-ion collisions.

\section{Hadron Resonance Gas Model}
\label{sec:hrg}
In the HRG model, we include all the relevant degrees of freedom of the confined, strongly 
interacting matter and also implicitly take into account the interactions that result in resonance 
formation~\cite{Karsch:2010ck}. It is well known that the fireball created in heavy ion collision 
does not remain static, rather expands both in longitudinal and transverse directions until freeze 
out occurs. However, to keep the model simple, we first consider a static homogeneous fireball and 
flow effects are included subsequently.

In heavy ion collision, no fluctuation would be seen in measurements with full phase space coverage 
as $B$, $Q$ and $S$ are strictly conserved. However, since most of the experiments cover only 
limited phase space, the part of the fireball accessible to the measurements may resemble with the 
Grand Canonical Ensemble (GCE) where energy (momentum), charge and number are not conserved 
locally. In general, the magnitude of multiplicity fluctuations and correlations in limited phase 
space crucially depends on the choice of the statistical ensemble that imposes different 
conservation laws \cite{hauer1,hauer2}. Since no extensive quantities like energy, momentum and 
charge are needed to be locally conserved 
in GCE, the particles following Maxwell-Boltzmann distribution are assumed to be 
uncorrelated and fluctuations are expected to follow Poisson statistics even in the limited phase 
space when quantum effects are ignored. In case of particles following Bose-Einstein or Fermi-Dirac 
distributions, within finite phase space, Poisson statistics is not expected to be obeyed and hence 
the deviations from Poisson limit can be studied.  

In the ambit of GCE framework , the logarithm of the partition function ($Z$) in the HRG model  
is given as,
\begin{eqnarray}
\label{eq:eq1}
\ln Z(T, V, \mu) = \sum_{B}\ln Z_{i}(T, V, \mu_i) + \sum_{M}\ln Z_{i}(T, V, \mu_i)\ ,
\end{eqnarray}
where,
\begin{eqnarray}
\ln Z_{i}(T, V, \mu_{i})= \pm\frac{Vg_{i}}{2\pi^2}\int 
d^3{p}\ln{\big\{1\pm\exp[(\mu_{i}-E)/T}]\big\},
\label{eq:eq2}
\end{eqnarray}
$T$ is the temperature, $V$ is the volume of the system, $\mu_{i}$ is the chemical potential, $E$ is
the energy and $g_{i}$ is the degeneracy factor of the $i^{th}$ particle. The total chemical 
potential $\mu_{i}$ = $B_{i}\mu_{B}$ + $Q_{i}\mu_{Q}$ + $S_{i}\mu_{S}$, where $B_{i}$, $Q_{i}$ and 
$S_{i}$ are the baryon, electric charge  and strangeness number of the $i^{th}$ particle,
with corresponding chemical potentials $\mu_{B}$, $\mu_{Q}$ and $\mu_{S}$, respectively. The '+' 
and '-' signs are for baryons and mesons respectively.
The thermodynamic pressure ($P$) can then be deduced for the limit of large volume as
\begin{equation}
\label{eq:eq3}
\frac{P}{T^4}=\frac{1}{VT^3} \ln{Z_{i}}= \pm\frac{g_{i}}{2\pi^2T^3}\int 
d^3{p}\ln{\big\{1\pm\exp[(\mu_{i}-E)/T}]\big \}.
\end{equation}

The $n^{th}$ order generalized susceptibility for baryons can be expressed as \cite{Karsch:2010ck},
 \begin{eqnarray}
 \label{eq:eq4}
 \chi_{x,baryon}^{(n)}=\frac{X^n}{VT^3}
\int{d^{3}p}\sum_{k=0}^{\infty}{(-1)^k}
(k+1)^n \\ \nonumber exp\bigg\{\frac{-(k+1)E } { T}\bigg\} {exp\bigg\{ \frac{(k+1)\mu}{T}\bigg\}},
\,
 \end{eqnarray}
and for mesons,
 \begin{eqnarray}
 \label{eq:eq5}  
 \chi_{x,meson}^{(n)}=\frac{X^n}{VT^3}
\int{d^{3}p}\sum_{k=0}^{\infty}(k+1)^n \\ \nonumber exp\bigg\{
\frac {-(k+1)E } { T}\bigg\} {exp\bigg\{ \frac{(k+1)\mu}{T}\bigg\}}.
\,
 \end{eqnarray}
The factor $X$ represents either $B$, $Q$ or $S$ of the $i^{th}$ particle depending on whether
the computed $\chi_{x}$ represents baryon or electric charge or strangeness susceptibility.

For a particle of mass $m$ in static fireball with $p_T$, $\eta$ and $\phi$ (azimuthal angle), the 
volume element ($d^{3}p$) and energy (E) can be written as 
$d^{3}p=p_T m_T \cosh\eta$$d{p_{T}}$$d\eta$$d\phi$
and $E$ = $m_T\cosh\eta$, where $m_{T}$=$\sqrt{p_{T}^{2} + m^{2}}$, respectively. 
The experimental acceptances can be incorporated by considering the appropriate integration ranges 
in $\eta$, $p_{T}$, $\phi$ and charge states by considering the values of $|X|$.
The total generalized susceptibilities will then be the sum of the contribution from baryons and mesons as, 
$\chi^{(n)}_x = \sum \chi^{(n)}_{x,baryon} + \sum \chi^{(n)}_{x,meson}$.

In order to make the connection with the experiments, the beam energy dependence of $\mu$ and $T$
parameters of the HRG model needs to be provided. These are extracted from a statistical thermal model 
description of measured particle yields in the experiment at various $\sqrt{s_{\rm {NN}}}$
~\cite{Cleymans:2005xv,Becattini1,Becattini2}. 
This is followed by the parameterization of $\mu_B$ and $T$ as a function of 
$\sqrt{s_{\rm {NN}}}$~\cite{Cleymans:2005xv}. The $\mu_B$  dependence of the temperature is 
given as $T(\mu_B) = a - b\mu_B^2 -c \mu_B^4$ with  $a =  0.166 \pm 0.002$ GeV, $b = 0.139 \pm 
0.016$ GeV$^{-1}$, and $c = 0.053 \pm 0.021$ GeV$^{-3}$. The $\sqrt{s_{\rm {NN}}}$ dependence of 
$\mu_B$ is given as $\mu_B(\sqrt{s_{NN}}) = \frac{d}{1 + e\sqrt{s_{NN}}}$
with $d = (1.308\pm 0.028)$ GeV and  $e = (0.273 \pm 0.008)$ GeV$^{-1}$. Further the 
ratio of baryon to strangeness chemical potential is parameterized as 
$\frac{\mu_S}{\mu_B} \simeq 0.164 + 0.018 \sqrt{s_{NN}}$. We have checked that the value of $T$ and 
$\mu_{B}$ obtained using the yields extrapolated to $4\pi$ or from mid-rapidity measurements, have 
little impact on our study. However in order to study the rapidity ($\eta$) dependence, the $\mu_B$ 
parameterizations  $\mu_{B} = 0.024 + 0.011 \eta^{2}$  and  $\mu_{B} = 0.237 + 0.011 \eta^{2}$ 
at \sqrts{200}~\cite{Cleymans:2008jw} and \sqrts{17.3}~\cite{Becattini3}, respectively are used in 
the calculations.

\section{Results}
\label{sec:results}

\subsection{Kinematic acceptance in $\eta$ and $p_{T}$}

  \begin{figure}[t]
    \includegraphics[width=\linewidth]{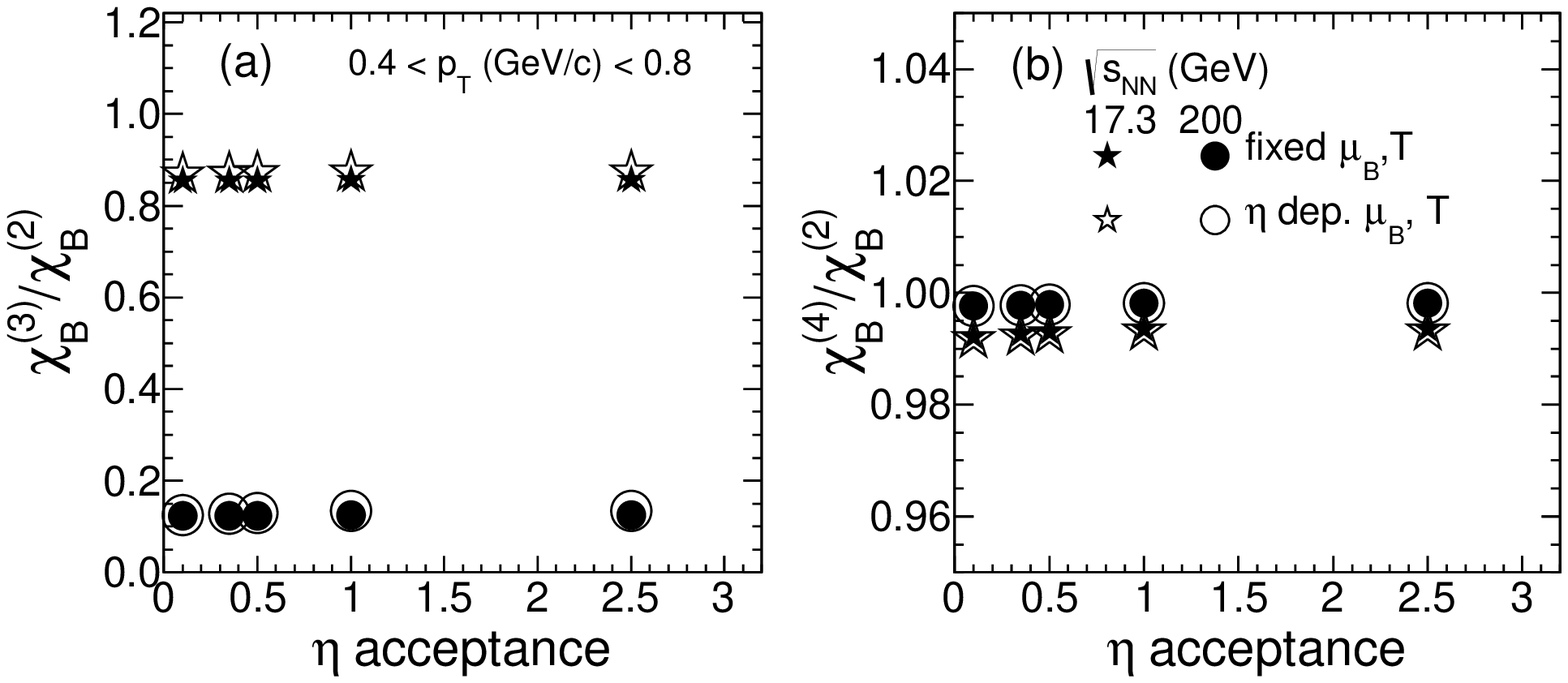}
    \includegraphics[width=\linewidth]{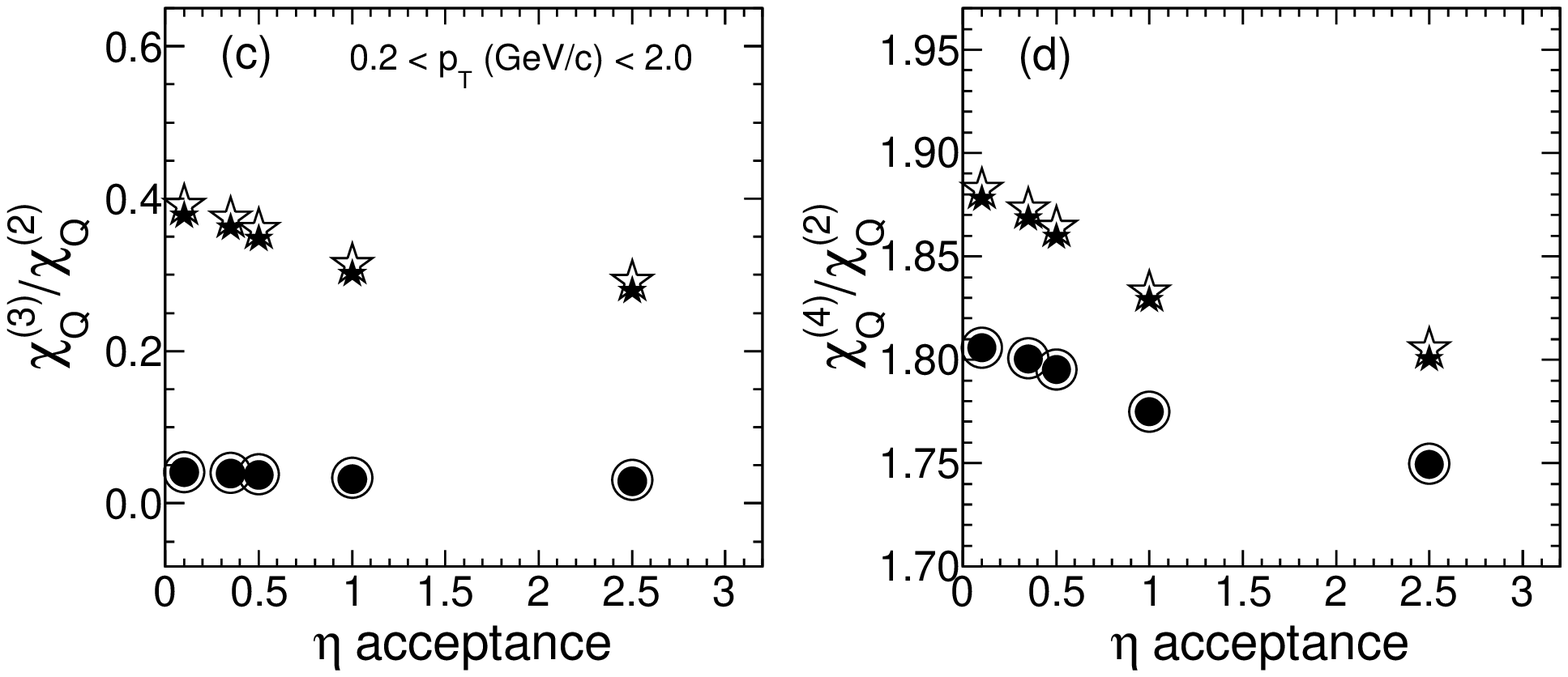}
    \includegraphics[width=\linewidth]{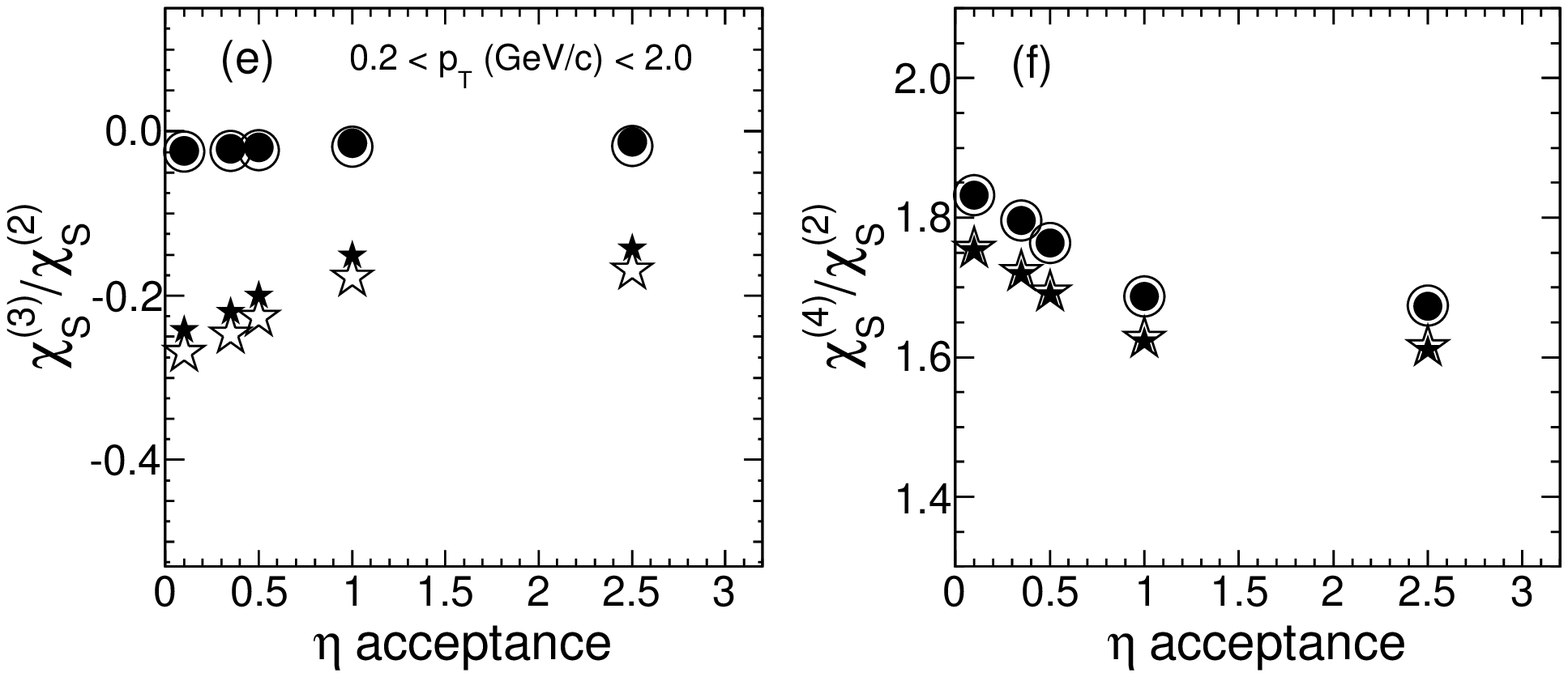}
    \caption{\label{fig:fig1}
      The $\eta$ acceptance dependence of $\chi_{x}^{(3)}/\chi_{x}^{(2)}$ and  
      $\chi_{x}^{(4)}/\chi_{x}^{(2)}$ for two different beam energies. In panel (a) and (b) $x$ = 
net-baryon $B$, (c) and (d)   $x$ = net-charge $Q$, and in (e) and (f) 
$x$ = net-strangeness $S$. Also shown are the results with (labeled ``dep.'') and without 
(labeled ``fixed'') the $\eta$  dependence of chemical freeze-out parameters $\mu_B$ and $T$. }
  \end{figure}
Figure~\ref{fig:fig1} shows the variation of $\chi_{x}^{(3)}/\chi_{x}^{(2)}$ and 
$\chi_{x}^{(4)}/\chi_{x}^{(2)}$ as a function of $\eta$ acceptance for \sqrts{200} and  
\sqrts{17.3}. Where $x$ stands for either net-baryon ($B$) (Fig.~\ref{fig:fig1} (a) and (b)), 
net-charge ($Q$) (Fig.~\ref{fig:fig1} (c) and (d)), or net-strangeness ($S$) (Fig.~\ref{fig:fig1} 
(e) and (f)). For each beam energy we show the effect of considering HRG parameters ($\mu$, $T$) 
fixed to parameterization based on mid-rapidity data and also a parameterization based on the 
$\eta$ dependent value of ($\mu$, $T$). 
The difference between the two cases are small. For the subsequent studies we only present results 
for different $\sqrt{s_{\rm {NN}}}$ using the parameterization of the chemical freeze-out parameters 
based on the measurement of particle yields at mid-rapidity. A clear dependence of  
$\chi_{x}^{(3)}/\chi_{x}^{(2)}$ and  $\chi_{x}^{(4)}/\chi_{x}^{(2)}$
on $\eta$ acceptance is observed for net-charge (Fig.~\ref{fig:fig1} (c) and (d)) 
and net-strangeness (Fig.~\ref{fig:fig1} (e) and (f)). The $\chi_{B}^{(3)}/\chi_{B}^{(2)}$ and  
$\chi_{B}^{(4)}/\chi_{B}^{(2)}$ values (Fig.~\ref{fig:fig1} (a) and (b)) are however found to be 
independent of $\eta$ acceptance for the two beam energies studied. This underscores the need to 
carefully consider $\eta$ acceptance effects when comparing HRG model results to experimental data, 
especially for net-charge and net-strangeness fluctuation measures.

  \begin{figure}[t]
    \includegraphics[width=\linewidth]{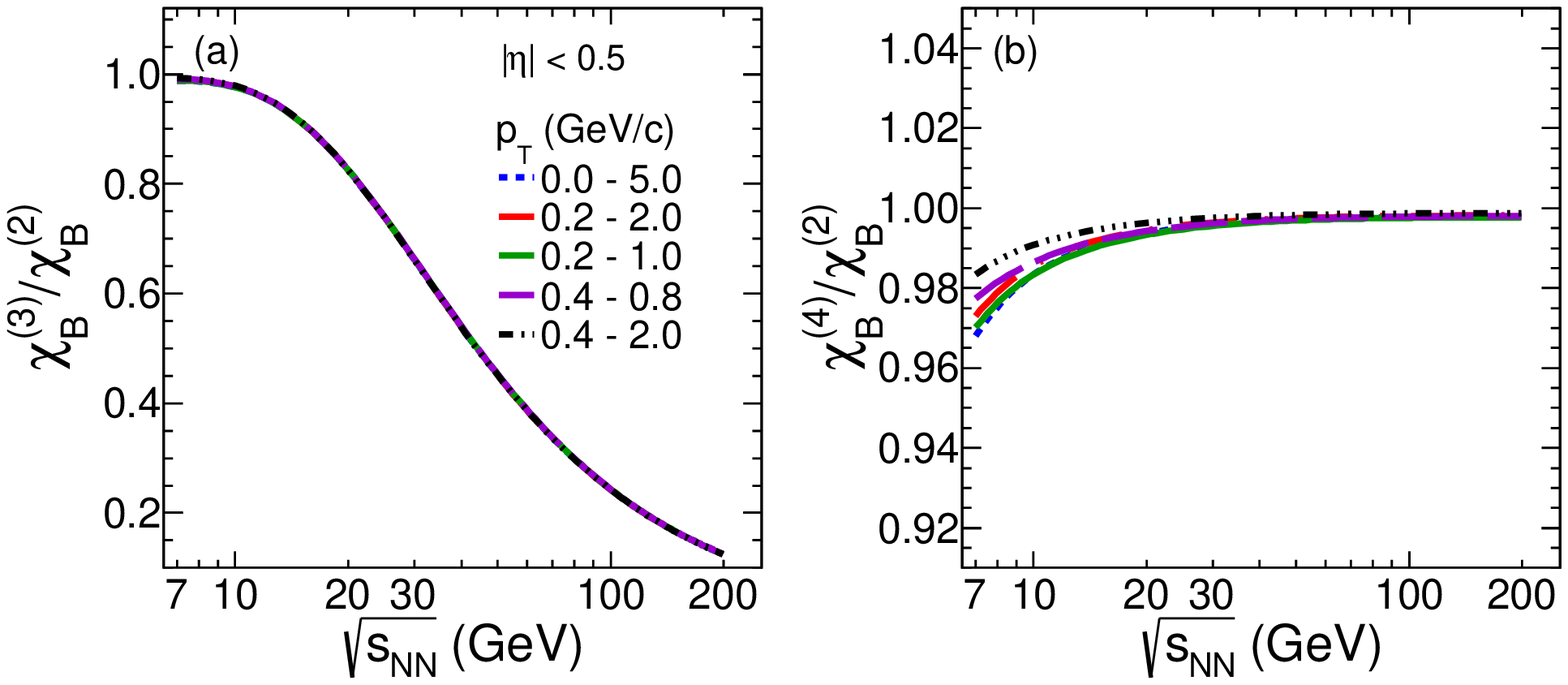}
    \includegraphics[width=\linewidth]{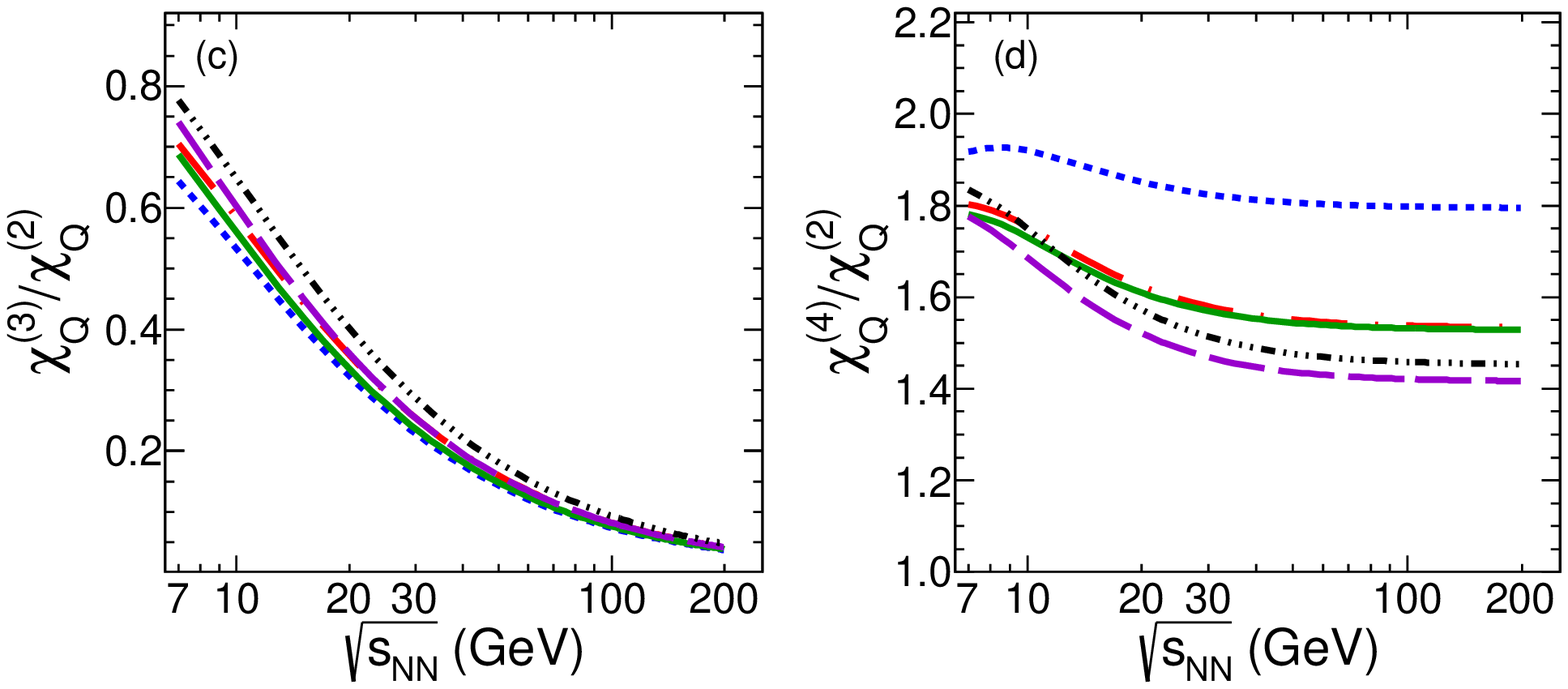}
    \includegraphics[width=\linewidth]{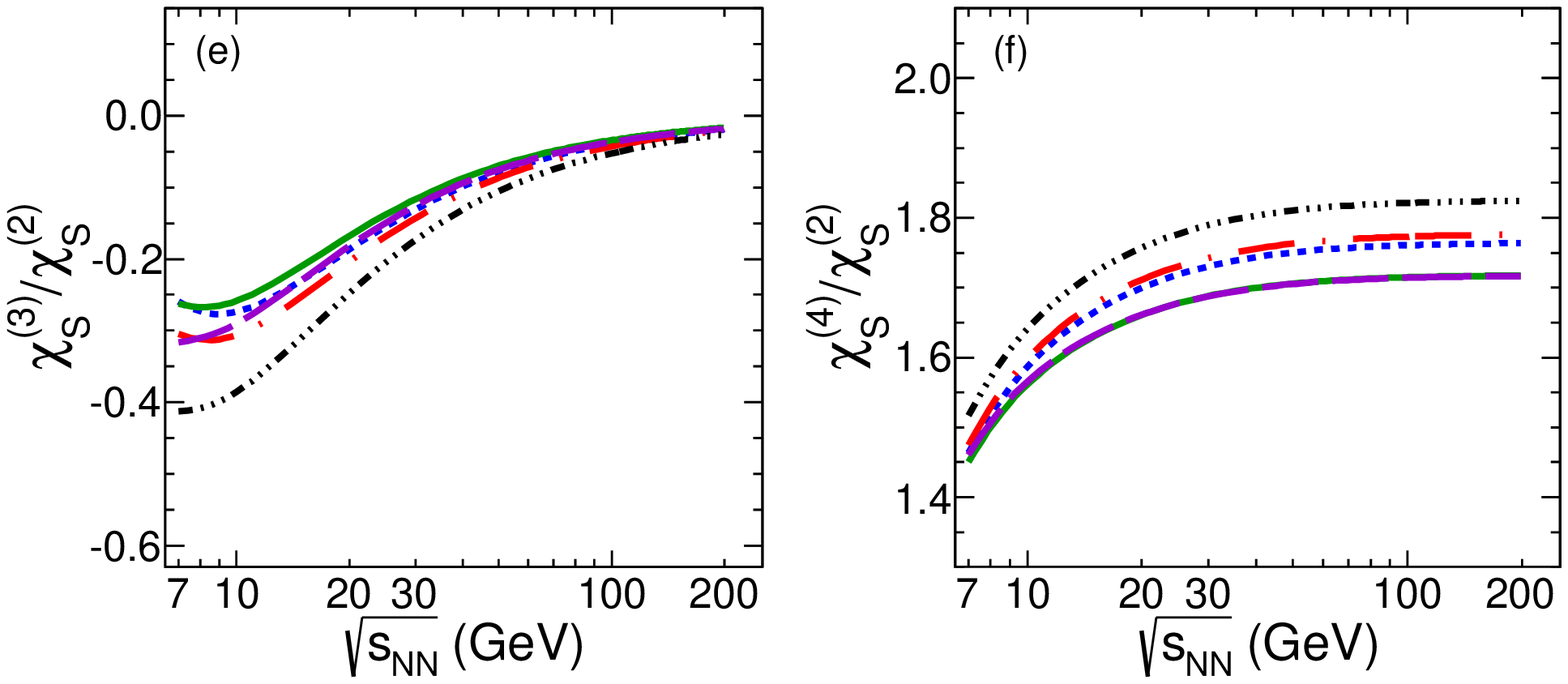}
    \caption{\label{fig:fig2}
      (Color online) The $p_T$ acceptance dependence of $\chi_{x}^{(3)}/\chi_{x}^{(2)}$ and  
      $\chi_{x}^{(4)}/\chi_{x}^{(2)}$ for different $\sqrt{s_{\rm {NN}}}$. Where $x$ stands for 
either net-baryon ($B$) (panels (a) and (b)), net-charge ($Q$) (panels (c) and (d)), and 
net-strangeness ($S$) (panels (e) and (f)).}
  \end{figure}

Figure~\ref{fig:fig2} shows the  variation of $\chi_{x}^{(3)}/\chi_{x}^{(2)}$ and  
$\chi_{x}^{(4)}/\chi_{x}^{(2)}$ as a function of $\sqrt{s_{\rm {NN}}}$ for various $p_{T}$ 
acceptances. The choice of these particular values of $p_{T}$ acceptance ranges are motivated by 
existence of the corresponding experimental measurements 
~\cite{Aggarwal:2010wy,McDonald:2012ts,Mitchell:2012mx}. 
It is observed that $\chi_{x}^{(3)}/\chi_{x}^{(2)}$ and  $\chi_{x}^{(4)}/\chi_{x}^{(2)}$ 
have a clear $p_{T}$ acceptance dependence at all beam energies for net-charge (Fig.~\ref{fig:fig2} 
(c) and (d)) and net-strangeness (Fig.~\ref{fig:fig2} (e) and (f)). However the $p_{T}$ acceptance 
dependences for net-baryon (Fig.~\ref{fig:fig2} (a) and (b)) is substantially weaker. Hence the 
$p_{T}$ acceptance study also emphasizes the need to consider the actual experimental acceptance for 
model comparisons in fluctuation measures. At the same time both the kinematic acceptance studies 
in $\eta$ and $p_{T}$ show net-baryon fluctuation measures are least affected.

\subsection{Conserved charge states}

  \begin{figure}[ht]
    \includegraphics[width=\linewidth]{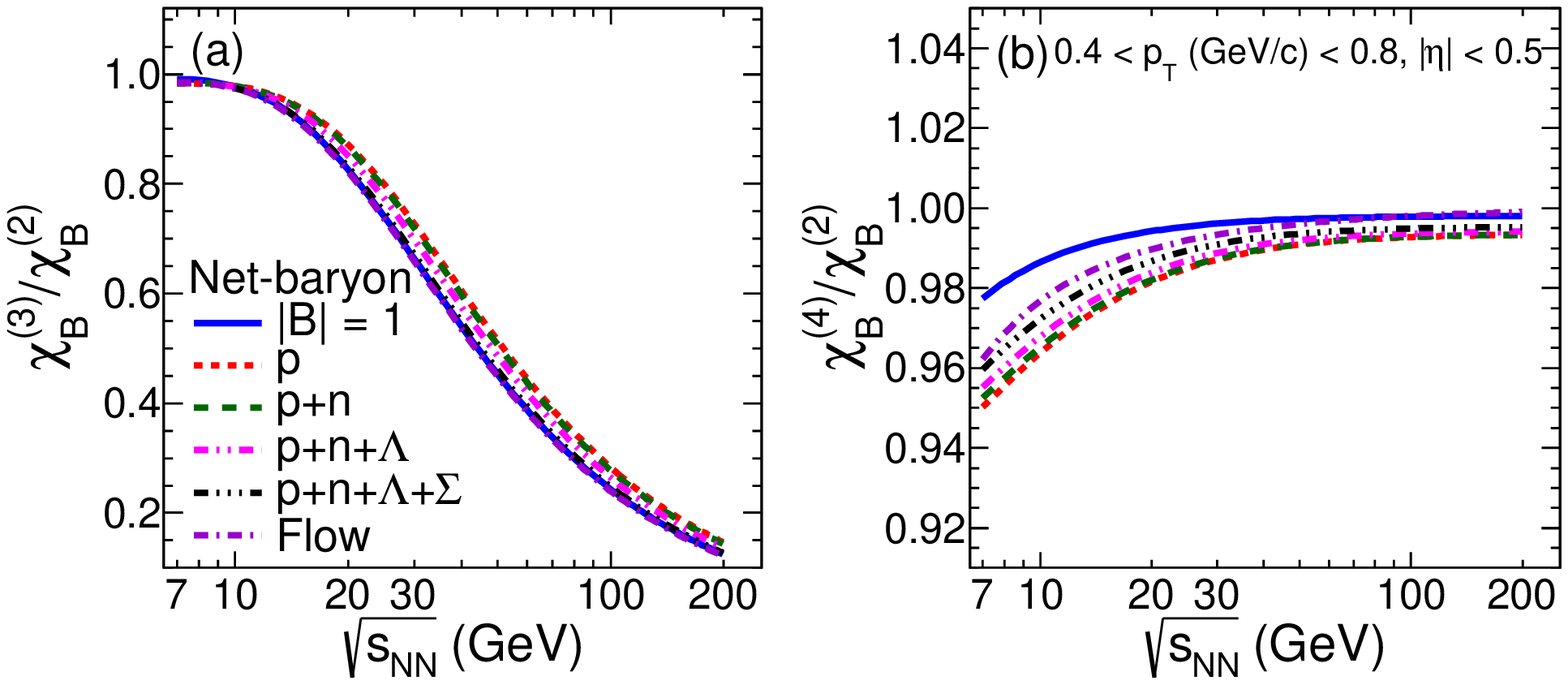}
    \includegraphics[width=\linewidth]{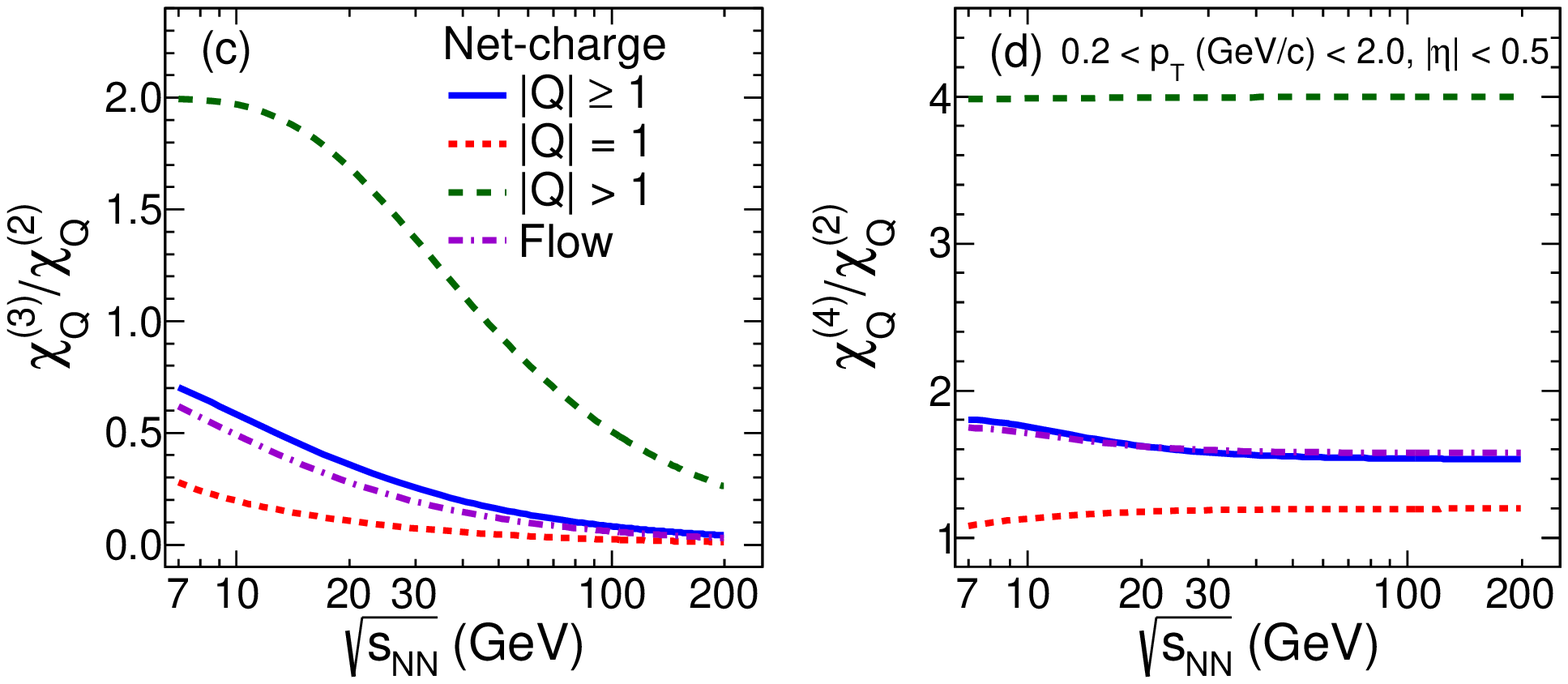}
    \includegraphics[width=\linewidth]{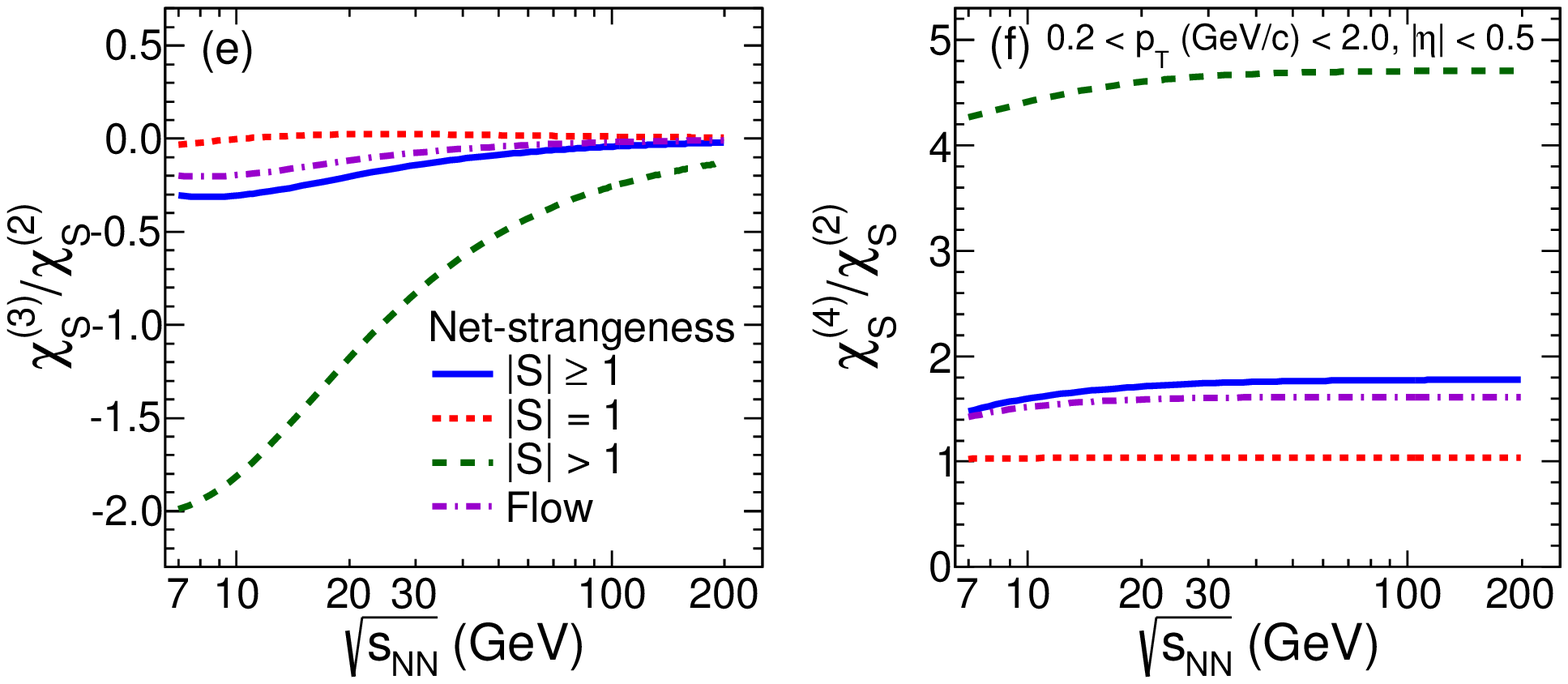}
    \caption{\label{fig:fig3}
      (Color online) The variation of $\chi_{x}^{(3)}/\chi_{x}^{(2)}$ and  $\chi_{x}^{(4)}/\chi_{x}^{(2)}$
for net-baryon ($B$), net-charge ($Q$), and net-strangeness ($S$) as a function of collision energy 
  ($\sqrt{s_{\rm {NN}}}$).The results are shown for different baryons (panels (a) and (b)), electric 
charge states (panels (c) and (d)) and strangeness number (panels (e) and (f)) considered in the 
calculation.  }
  \end{figure}
Figure~\ref{fig:fig3} shows the  variation of $\chi_{x}^{(3)}/\chi_{x}^{(2)}$ and  
$\chi_{x}^{(4)}/\chi_{x}^{(2)}$ as a function of $\sqrt{s_{\rm {NN}}}$ for various types of baryons 
(Fig.~\ref{fig:fig3} (a) and (b)), values of electric charge states, $|Q|$ = 1 and $|Q|$ $>$ 1 
(Fig.~\ref{fig:fig3} (c) and (d)), and values of strangeness number, $|S|$ = 1 and $|S|$ $>$ 1 
(Fig.~\ref{fig:fig3} (e) and (f)). For each of the cases the observables are compared to the 
respective values  with inclusion of all conserved charge states and baryons. We find a strong 
dependence of the $\chi_{Q}^{(3)}/\chi_{Q}^{(2)}$ and  $\chi_{Q}^{(4)}/\chi_{Q}^{(2)}$ on whether we 
consider $|Q|$ = 1 or $|Q|$ $>$ 1, both differing from the case of inclusion of all charge states. 
Same is the situation for net-strangeness. On the other hand, successive inclusion of different 
baryons, starting with protons seems to have some small effect on the 
$\chi_{B}^{(3)}/\chi_{B}^{(2)}$ and  $\chi_{B}^{(4)}/\chi_{B}^{(2)}$ values only at the lower beam 
energies.The absence of baryons with $|B|$ $>$ 1 makes the net-baryon number fluctuations more 
advantageous and less prone to kinematic acceptances as compared to the net-charge or 
net-strangeness fluctuation measures.

\subsection{Effect of flow}
To study the effect of flow on the ratios of susceptibilities in the HRG model, we now consider an 
expanding fireball with four velocity,
\begin{equation}
\label{eq:eq6}
 u^{\mu} = \mathrm{cosh}y_{T}(\mathrm{cosh}y_0, \mathrm{tanh}y_{T}, 0, \mathrm{sinh}y_0),
\end{equation}
where $y_0$ is the longitudinal and $y_T$ = $\mathrm{tanh}^{-1}$($\beta_r$) is the transverse 
rapidity of the fireball flowing with radial flow velocity $\beta_r$. 
The four-momentum of the particle can be expressed as:
\begin{equation}
\label{eq:eq7}
 p^{\mu} = (m_{T} \mathrm{cosh}\eta, p_T\mathrm{cos}\phi, 
p_T\mathrm{sin}\phi, m_{T}\mathrm{sinh}\eta)
\end{equation}
In the presence of flow, the logarithm of the partition function for $i^{th}$ particle having 
four 
momentum $p^\mu$ can be expressed as,
\begin{eqnarray}
 \ln Z_{i}(T, V, \mu_i) = \pm\frac{g_{i}}{{(2\pi)}^3} 
\int_\sigma d^3{p}\frac{ p^\mu d\sigma_\mu}{p^0} \\ \nonumber 
 \ln{\big\{1\pm\exp[(\mu_{i}-p^\mu u_\mu)/T]}\big\},
\label{eq:eq8}
\end{eqnarray}
where $\sigma$ represents the space-time surface whose surface elements can be 
represented by four vector denoted by $d\sigma_\mu$ and $p^{0}$ is the energy of the 
particle. Assuming instantaneous freeze out (at time $\tau_f$) in the radial direction 
$r$, $p^\mu d\sigma_\mu$ simplifies as~\cite{rus},
\begin{eqnarray}
\label{eq:eq9}
p^\mu d\sigma_\mu=\tau_f r dr d\phi dy_0 m_T \mathrm{cosh}(\eta-y_0).
\end{eqnarray} 
The limit of integration for $r$ varies from $0$ to $R_f$ (freeze out radius), $\phi$ from $0$ 
to $2\pi$, $y_0$ from $-y_0^{min}$ to $y_0^{max}$ (= $ln (\sqrt{s_{NN}}/m_p)$, where $m_p$ is mass 
of proton). The other variables $p_T$ and $\eta$ vary within the experimental acceptances.
Note that we recover eq. \ref{eq:eq2} for the case of static fireball ($y_T=y_0=0$) where $p^\mu 
u_\mu=E$ and the integral over $(p^\mu d\sigma_\mu)/p^0$ becomes proportional to $4\pi V$. 
Therefore, for a constant $\beta_r$, the flow effect on the susceptibilities can be incorporated  
by replacing $d^3p$ integral in eq. \ref{eq:eq4} and eq. \ref{eq:eq5} with $p_T dp_T d\eta d\phi 
dy_0 m_T \mathrm{cosh}(\eta-y_0)$ and the energy $E$ in the exponentials by the invariant $p^\mu 
u_u$ as defined above. 
Further, under the assumption that the flow velocity $\beta_r$ is independent of radial position,
the $r$ integration turns out to be a constant which is proportional to the volume at freeze-out.
For the simplicity of the calculations, we have used a constant $\beta_r$ (same as $\beta_s$ of 
ref~\cite{star_prc}). 
Figure~\ref{fig:fig3} also shows the effect of flow (longitudinal + transverse) on the ratios of 
susceptibilities as a function of collision energy. 
It is noticed that
the ratios of the susceptibilities like $\chi_{x}^{(4)}/\chi_{x}^{(2)}$ for net-baryon, net-charge 
and net-strangeness are affected by less than (2 - 4)$\%$ as compared to the corresponding static 
values represented by solid blue line. 
 
  \begin{figure}[ht]
    \includegraphics[width=\linewidth]{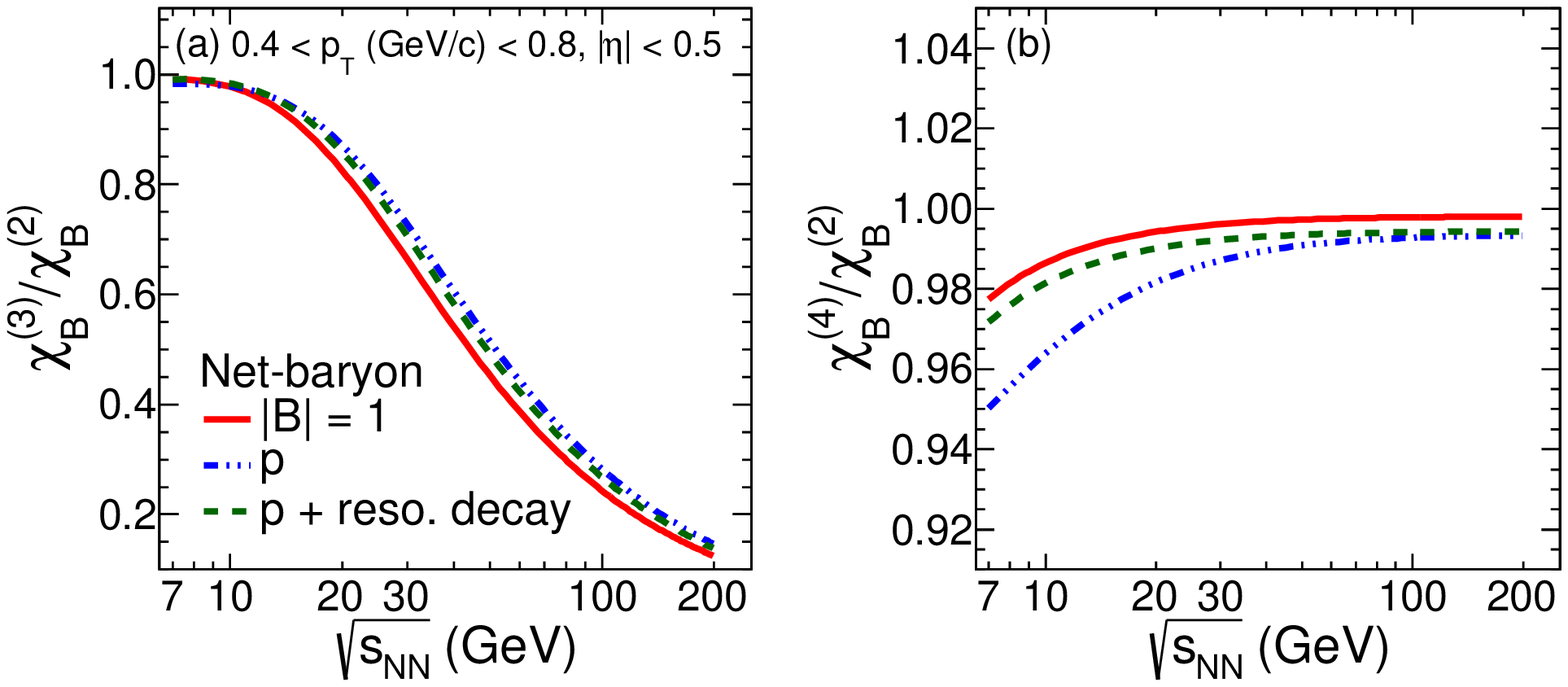}
    \includegraphics[width=\linewidth]{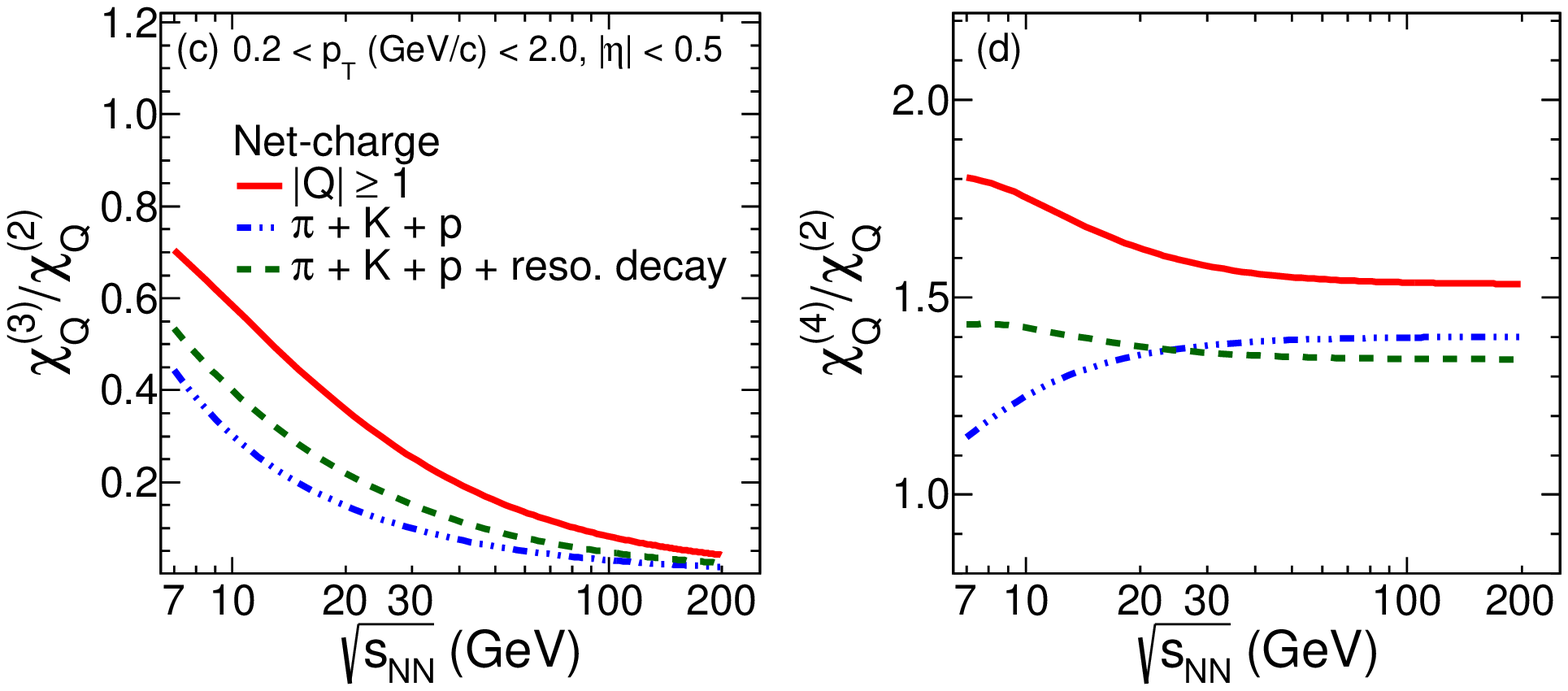}
    \includegraphics[width=\linewidth]{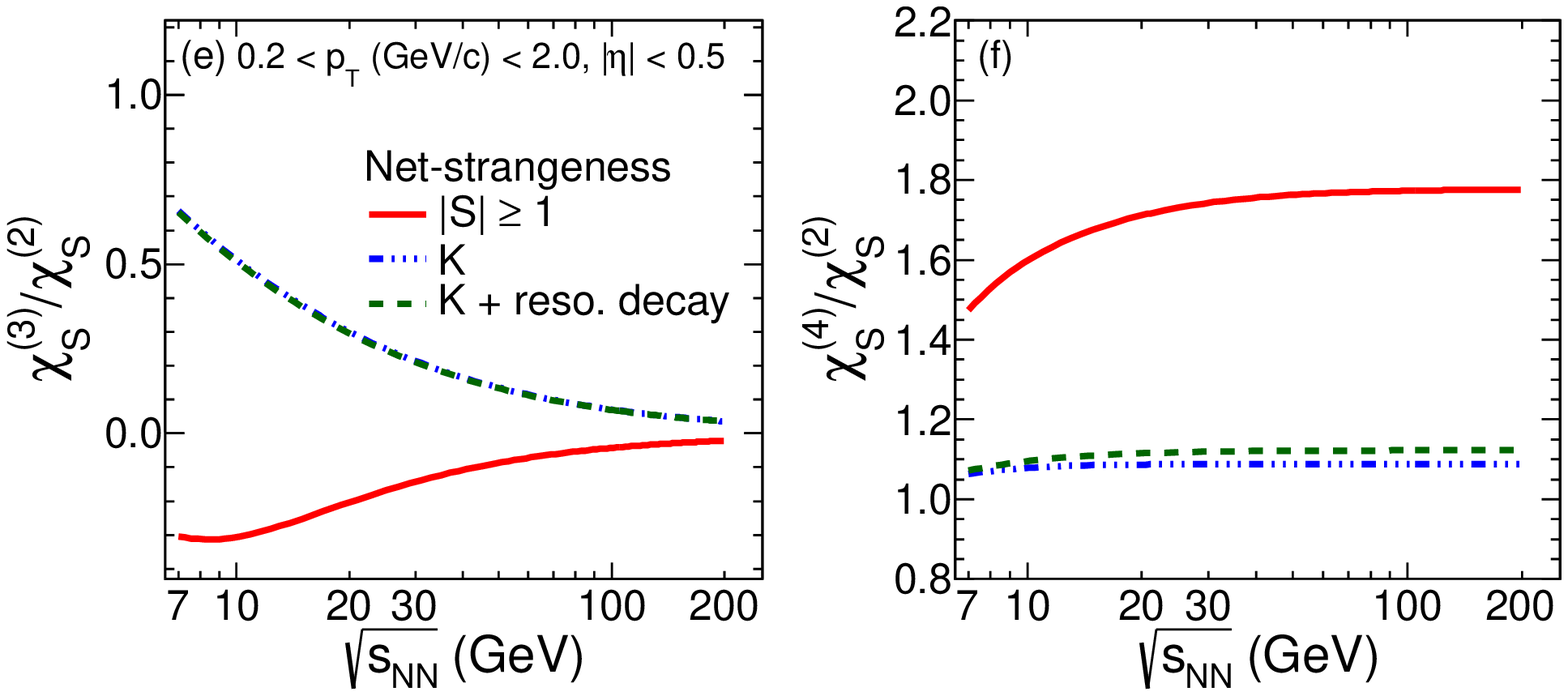}
    \caption{\label{fig:fig4} (Color online) The variation of $\chi_{x}^{(3)}/\chi_{x}^{(2)}$ 
    and  $\chi_{x}^{(4)}/\chi_{x}^{(2)}$ for net-baryon ($B$), net-charge ($Q$), and 
net-strangeness ($S$)  with different beam energies ($\sqrt{s_{\rm {NN}}}$) with and without 
resonance decay daughter particle acceptance effects. }
  \end{figure}
  
\subsection{Resonance decay}

The generalized susceptibility of $n^{th}$ order can be written as, 
\begin{eqnarray}
\chi^n=\sum_p \chi^n_p+ 
\sum_R \epsilon^n_R \chi^n_R,
\end{eqnarray}
 where $\chi^n_p$  and $\chi^n_R$ are the  
contributions to the $n^{th}$ order susceptibility due to primordial and resonance yields 
respectively.
The factor $\epsilon^n_R$ is an 
event averaged efficiency at which resonance $R$ contributes to the generalized 
susceptibility. 
Note that $\epsilon^n_R=1$ when all the decay particles are fully accepted
and $\epsilon^n_R<1$ due to finite detector and kinematic acceptances. 
Consider the example of the resonance $\Delta^{++}$ which decays into $p$ and $\pi^+$ with branching 
ratio $b$. Using a toy Monte Carlo simulation, we generate $\Delta^{++}$ distribution with Poisson 
statistics and build the charge distributions both for $\Delta^{++}$ and the decay particles 
(proton and pion together) within the experimental acceptances. 
The Poisson distribution for $\Delta^{++}$ is a reasonable assumption
as momentum distributions of resonance particles can be approximated by the classical 
Maxwell-Boltzmann function
due to their large masses, although it is not true for their decay  products. Therefore, we estimate the
efficiency $\epsilon^n$ by taking the ratios of the $n^{th}$ order cumulant of the charge 
distributions
after and before decays. Similar procedures are adopted for other resonances to estimate
the average efficiencies 
$\epsilon^n_R$ depending on the charge, strangeness and baryon number as appropriate.

The results of $\chi_{x}^{(3)}/\chi_{x}^{(2)}$ and  
$\chi_{x}^{(4)}/\chi_{x}^{(2)}$ with and without considering the effect of resonance 
daughter particle acceptances are shown in Fig.~\ref{fig:fig4}. 
In Fig.~\ref{fig:fig4}(a) and (b), three cases for $\chi_{B}^{(3)}/\chi_{B}^{(2)}$ and 
$\chi_{B}^{(4)}/\chi_{B}^{(2)}$ are shown within a realistic acceptance of $|\eta|$ $<$ 0.5 
and 0.4 $<$ $p_{T}$ $<$ 0.8 GeV/$c$. The results for all baryons without any resonance decay 
(solid red curve), results for protons without any resonance decay contribution (dotted blue curve) 
and results for protons with resonance decay (dashed green curve). Similarly in Fig. 
~\ref{fig:fig4}(c) and (d) shows $\chi_{Q}^{(3)}/\chi_{Q}^{(2)}$ and $\chi_{Q}^{(4)}/\chi_{Q}^{(2)}$ 
respectively, for all charges without resonance decay (solid red curve), pions, kaons and protons 
without resonance decay (dotted blue curve) and  pions, kaons and protons with resonance decay 
(dashed green curve). Fig.~\ref{fig:fig4}(e) and (f), are shown the results for all strangeness 
without resonance decay (solid red curve), kaons without resonance decay (dotted blue curve) and 
kaons with resonance decay (dashed green curve). 

It may be mentioned here that the efficiency $\epsilon^n_R$ is an event averaged quantity and will have
fluctuations on an event by event basis. Although $\epsilon^n_R$ has inherent fluctuation,
a rough estimate shows that it's effect on the ratio of $\chi_{Q}^{(4)}/\chi_{Q}^{(2)}$ is  
$\sim$$2\%$ which we have ignored in the present study. Therefore, the present estimate of the 
effect of resonance decays on the ratio of susceptibilities are approximate. 
Nevertheless, it brings out the importance of resonance decays
which certainly affect all the ratios. With these assumption, the effects of resonance decay due to 
finite experimental acceptances are large for net-charge and net-strangeness as compared to 
net-baryons. The acceptance used in this study are modest and are close to the present 
experimental acceptances. 


Through our work we have emphasized the need for considering experimental acceptances of various 
kinds in model, such as HRG, before they are considered to provide the baseline to experimental 
measurements for drawing physics conclusions. Hence in the 
Tables~\ref{tab:netchstar},~\ref{tab:netchphx},~\ref{tab:netkaon} and ~\ref{tab:netproton} we 
provide values of $\chi_{x}^{(3)}/\chi_{x}^{(2)}$ and $\chi_{x}^{(4)}/\chi_{x}^{(2)}$ for typical
ongoing experimental acceptances. The values quoted in the tables are for static fireball and 
without including the resonance decay products. The 
$\chi_{Q}^{(3)}/\chi_{Q}^{(2)}$ and $\chi_{Q}^{(4)}/\chi_{Q}^{(2)}$
values are provided for two typical acceptances $|\eta|$ $<$ 0.5, 0.2 $<$ $p_{T}$ $<$ 2.0 GeV/$c$ 
and $|\eta|$ $<$ 0.35, 0.3 $<$ $p_{T}$ $<$ 1.0 GeV/$c$ (Table~\ref{tab:netchstar} and 
Table~\ref{tab:netchphx}). The $\chi_{S}^{(3)}/\chi_{S}^{(2)}$ and $\chi_{S}^{(4)}/\chi_{S}^{(2)}$ 
are provided for a typical acceptance of $|\eta|$ $<$ 0.5, 0.2 $<$ $p_{T}$ $<$ 2.0 
GeV/$c$~(Table~\ref{tab:netkaon}). The $\chi_{B}^{(3)}/\chi_{B}^{(2)}$ and 
$\chi_{B}^{(4)}/\chi_{B}^{(2)}$ are provided for a typical acceptance of $|\eta|$ $<$ 0.5, 0.4 $<$ 
$p_{T}$ $<$ 0.8 GeV/$c$~(Table~\ref{tab:netproton}).

\begin{table}[ht]
\begin{center}
\caption{Ratios of the moments for net-charge within $|\eta|$ $<$ 0.5, and 
  0.2 $<$ $p_{T}$ $<$ 2.0 GeV/$c$.}

\begin{tabular}{|c|c|c|c|}
\hline
$\sqrt{s_{NN}}$ (GeV)& $\chi_{Q}^{(3)}/\chi_{Q}^{(2)}$ & $\chi_{Q}^{(4)}/\chi_{Q}^{(2)}$ \\[2pt]
\hline
      5        &  0.526  &    1.413  \\
      7.7      &  0.414  &    1.430  \\
      11.5     &  0.321  &    1.407  \\
      15       &  0.265  &    1.390   \\
      19.6     &  0.215  &    1.375  \\
      27       &  0.165  &    1.361  \\
      39       &  0.119  &    1.352  \\
      62.4     &  0.077  &    1.346  \\
      130      &  0.038  &    1.343   \\
      200      &  0.025  &    1.342  \\
      2760     &  0.002  &    1.341   \\ 
\hline
\end{tabular}
\label{tab:netchstar}
\end{center}
\end{table}

\begin{table}[ht]
\begin{center}
\caption{Ratios of the moments for net-charge within $|\eta|$ $<$ 0.35, and 
  0.3 $<$ $p_{T}$ $<$ 1.0 GeV/$c$.}
\begin{tabular}{|c|c|c|c|}
\hline
$\sqrt{s_{NN}}$ (GeV) & $\chi_{Q}^{(3)}/\chi_{Q}^{(2)}$ & $\chi_{Q}^{(4)}/\chi_{Q}^{(2)}$ \\[2pt]
\hline
      5        &  0.432  &     1.231\\
      7.7      &  0.332  &     1.245\\
      11.5     &  0.256  &     1.233\\
      15       &  0.212  &     1.222 \\
      19.6     &  0.172  &     1.213\\
      27       &  0.132  &     1.205\\
      39       &  0.095  &     1.199\\
      62.4     &  0.062  &     1.196\\
      130      &  0.031  &     1.194\\
      200      &  0.020  &     1.193\\
      2760     &  0.001  &     1.193\\ 
\hline
\end{tabular}
\label{tab:netchphx}
\end{center}
\end{table}

\begin{table}[ht]
\begin{center}
\caption{Ratios of the moments for net-kaon within $|\eta|$ $<$ 0.5, and 
  0.2 $<$ $p_{T}$ $<$ 2.0 GeV/$c$.}
\begin{tabular}{|c|c|c|c|}
\hline
$\sqrt{s_{NN}}$ (GeV) & $\chi_{K}^{(3)}/\chi_{K}^{(2)}$ & $\chi_{K}^{(4)}/\chi_{K}^{(2)}$ \\[2pt]
\hline
5       &  0.726  &   1.058 \\
7.7     &  0.560  &   1.090 \\
11.5    &  0.428  &   1.108 \\
15      &  0.353  &   1.115 \\
19.6    &  0.288  &   1.120 \\
27      &  0.222  &   1.123 \\
39      &  0.162  &   1.125  \\
62.4    &  0.107  &   1.127 \\
130     &  0.054  &   1.128  \\
200     &  0.035  &   1.128 \\
2760    &  0.003  &   1.128\\
\hline
\end{tabular}

\label{tab:netkaon}
\end{center}
\end{table}

\begin{table}[ht]
\begin{center}
\caption{Ratios of the moments for net-proton within $|\eta|$ $<$ 0.5, and 
  0.4 $<$ $p_{T}$ $<$ 0.8 GeV/$c$.}
\begin{tabular}{|c|c|c|c|}
\hline
$\sqrt{s_{NN}}$ (GeV) & $\chi_{p}^{(3)}/\chi_{p}^{(2)}$ & $\chi_{p}^{(4)}/\chi_{p}^{(2)}$ \\[2pt]
\hline

5    &   0.981   &   0.961 \\
7.7  &   0.987   &   0.975 \\
11.5 &   0.962   &   0.984 \\
15   &   0.920   &   0.988 \\
19.6 &   0.851   &   0.991 \\
27   &   0.737   &   0.993 \\
39   &   0.586   &   0.995 \\
62.4 &   0.407   &   0.996 \\
130  &   0.210   &   0.997 \\
200  &   0.139   &   0.997 \\
2760 &   0.010   &   0.997 \\ 
\hline
\end{tabular}

\label{tab:netproton}
\end{center}
\end{table}

\section{Summary}
\label{sec:summary}
In summary, using a hadron resonance gas model we have studied the effect of limited experimental
acceptance on observables like $n^{th}$ order susceptibilities $\chi_{x}^{(n)}$, associated with 
conserved quantities like net-charge ($x$ = $Q$), net-strangeness ($x$ = $S$) and  net-baryon 
number ($x$ = $B$). The various order susceptibilities which can also be calculated in QCD based 
models are related to the moments ($\sigma$, $S$ and $\kappa$) of the corresponding measured 
conserved number distributions. These observables have been widely used to understand the freeze-out 
conditions in heavy-ion collisions and various aspects  of the phase structure of  the QCD phase 
diagram. Our study demonstrates the importance of considering experimental acceptances of different 
kinds before measurements are compared to theoretical calculations, specifically in the use of HRG 
model as a baseline for such fluctuation based study. We observe finite kinematic acceptances in 
$\eta$ and $p_{T}$ have a strong effect on the $\chi_{Q}^{(n)}$ and $\chi_{S}^{(n)}$ values. These 
susceptibilities are also very sensitive to the accepted electric charge states and strangeness 
states in the experiment. However, the effect of flow is less than (2 - 4)$\%$ on the ratios 
of susceptibilities for net-baryon, net-charge and net-strangeness and the improvements can be done 
in implementation of the radial dependent transverse flow velocities.
In addition, we find in this model that, the $\chi_{Q}^{(n)}$ and $\chi_{S}^{(n)}$ values depends 
on the experimental acceptance of the decay daughters from various resonances produced in high 
energy heavy-ion collisions. Within this model and the kinematic regions used in our study, we find 
that the dependence on acceptance and resonance decays are stronger for both net-charge and 
net-strangeness compared to that of net-baryons.

\noindent {\bf Acknowledgments }
We thank Sourendu Gupta for useful discussions related to this paper. BM is supported by the DST 
SwarnaJayanti project fellowship. PG acknowledges the financial support 
from CSIR, New Delhi, India.




\end{document}